\documentclass[twocolumn, amsmath,aps,prb,showpacs]{revtex4}
\usepackage{graphicx}

\begin{document}

\title{Microwave properties of DyBa$_2$Cu$_3$O$_{7-x}$ monodomains and related compounds in magnetic fields}

\author{N. Pompeo}
\affiliation{Dipartimento di Fisica ``E. Amaldi'' and Unit\`a CNISM, Universit\`a Roma Tre, Via della Vasca Navale 84, I-00146 Roma, Italy}
\author{E. Silva$^{a)}$ \footnote[0]{$^{a)}$ Corresponding author. E-mail: silva@fis.uniroma3.it}}
\affiliation{Dipartimento di Fisica ``E. Amaldi'' and Unit\`a CNISM, Universit\`a Roma Tre, Via della Vasca Navale 84, I-00146 Roma, Italy}
\author{M. Ausloos}
\affiliation{SUPRATECS,  B5a, B-4000 Li\`ege, Euroland}
\author{R. Cloots}
\affiliation{SUPRATECS, Dept. of Chemistry (B6), University of Li\`ege, B-4000 Li\`ege, Belgium}

\begin{abstract}
\noindent We present  a microwave characterization of a DyBa$_{2}$Cu$_{3}$O$_{7-x}$ single domain, grown by the top-seeded melt-textured technique.
We report  the (a,b) plane field-induced surface resistance, $\Delta R_s(H)$, at 48.3 GHz, measured by means of a cylindrical metal cavity in the end-wall-replacement configuration. Changes in the 
cavity quality factor $Q$ against the applied magnetic field yield  $\Delta R_s(H)$ at fixed temperatures. The temperature range [70 K ; $T_c$] was explored. The magnetic field $\mu_0 H <$ 0.8 T was applied along the $c$ axis. The field dependence of $\Delta R_s(H)$ does not exhibit the steep, step-like increase at low fields typical of weak-links. This result  indicates the single-domain character of the sample under investigation. $\Delta R_s(H)$ exhibits  a nearly square-root dependence on $H$, as expected for fluxon motion. From the analysis of the data in terms of motion of Abrikosov vortices we estimate the temperature dependences of the London penetration depth $\lambda$ and the vortex viscosity $\eta$, and their zero-temperature values $\lambda(0)=$165 nm and $\eta(0)=$ 3 10$^{-7}$ Nsm$^{-2}$, which are found in excellent agreement with reported data in YBa$_{2}$Cu$_{3}$O$_{7-x}$ single crystals. 
Comparison of microwave properties with those of related samples indicate the need for reporting data as a function of $T/T_c$  in order to obtain universal laws.
\end{abstract}

\pacs{74.72.Jt, 74.25.Nf, 74.25.Qt}

\maketitle

\section{Introduction}
Microwave frequencies might be one of the best physical regions for devices, typically in the telecommunication field, for High critical Temperature ($T_{c}$) Superconductors (HTcS), including ceramics. \cite{ma1,ma2,ma3} Moreover the microwave response of  HTcS provides important information about fundamental physics \cite{narlikar} besides the essential issues for technological applications \cite{gallop}. Several reviews of microwave properties of such superconductors can be found in Refs.(\onlinecite{narlikar,gallop,ma4,ma5,ma6,golos,silvaNova}). It has been already observed \cite{mausloos} that data are contradictory and even paradoxical. It has been questioned whether observed effects were of intrinsic or extrinsic origin, \cite{mausloos,nedkov} e.g. are due to the order parameter symmetry or weak links.\\
One way to solve such puzzles is to work on single crystals, but it is technically very hard to obtain these with an appropriate dimension. Thin films have been much studied. \cite{golos,silvaNova,others} Monodomains have not received any comparable attention from the point of view of microwave properties. This class of materials is of considerable interest for applications and, once the absence of extrinsic effects is fully assessed, they can prove extremely interesting for fundamental studies.\\
Our aim is to investigate {\it (i)} to what extent the monodomains can be usefully studied as pure materials,  {\it (ii)}  whether they are as good as single crystal materials and  {\it (iii)}  to compare the microwave response of DyBa$_{2}$Cu$_{3}$O$_{7-x}$ (DyBCO) monodomains to well established properties of its parent compound, YBa$_{2}$Cu$_{3}$O$_{7-x}$  (YBCO).\\
At microwave frequencies the main physical quantity directly measurable in the experiment is the effective surface impedance which, in the case of bulk materials (thickness $t_{s}\gg \min(\lambda, \delta)$ where $\lambda$ and $\delta$ are the London penetration depth and skin depth, respectively), is given in the local limit by the usual expression:
\begin{equation}
\label{zbulk}
  Z_{s}(H,T)=R_{s}+{\rm i}X_{s}=
  \sqrt{ {\rm i}\omega\mu_{0}\tilde{\rho}}={\rm i}\omega\mu_0\tilde{\lambda}(H,T)
\end{equation}
\noindent where $\omega=2\pi\nu$ is the angular frequency of the electromagnetic (em) field, $\tilde{\rho}$ is the superconductor complex resistivity, $R_{s}$ and $X_{s}$ are the surface resistance and reactance and account for dissipation and electromagnetic energy storage, respectively, and the last equality introduces the complex penetration depth. \cite{cc} This expression has been widely used as the starting point in the analysis of the field-dependence of the surface impedance in superconductors 
(see, e.g., \cite{golos} and references therein). In this paper we concentrate on the field-induced dissipation, given by the change of the surface resistance upon application of a magnetic field, $\Delta R_{s}(H,T)=R_{s}(H,T)-R_{s}(0,T)$.\\
The paper is organized as follows. In Sect. II, we  briefly outline the monodomain growth and the microwave experimental technique. The microwave data and a discussion are found in Sect. III. Some conclusions are drawn in Sect. IV.
\section{Experimental section}
DyBCO single domains were prepared with precursor powders DyBa$_{2}$Cu$_{3}$O$_{7-x}$ and Dy$_{2}$BaCuO$_{5}$, produced by solid-state synthesis from Dy$_{2}$O$_{3}$, BaCO$_{3}$ and CuO powders. The powder mixture was pressed uni-axially to give cylindrical pellets of 10.8 mm diameter, which were melt-textured in atmospheric air conditions using a Nd-123 single-crystal seed. Large Ôquasi-single-crystalsÕ, mainly c-axis oriented, have been obtained, with $T_c\sim$ 88-89 K. Two samples were cut from the same pellet: sample A (thickness $\sim$~1mm) is the subject of the present study of the microwave properties, and exhibited an inflection point of the microwave resistance close to 88 K, consistent with $T_c$ as given by magnetization and dc resistivity data on similar samples.  Sample B has been characterized by magneto-optics to assess its quasi-single-crystal properties. \cite{magnetoopt} Magneto-optical pictures recorded at different fields showed that it has a global single domain aspect, with a very few cracks of large dimension and few, small inhomogeneities. Detailed high-frequency dissipation measurements were taken on sample A, appearing also to be a DyBa$_2$Cu$_3$O$_{7-x}$ (DyBCO) single domain, and are reported and discussed in Sect. III.\\
The magnetic field dependence of the microwave surface resistance of the DyBCO sample was measured by the cavity method in the field range 0.8 T $> \mu_0 H >$0. An extensive description of the experimental setup has been given in a separate publication, \cite{silva} whence we recall here only the basics of the measurement technique. The sample under study occupies an end wall of a cylindrical (diameter = 8.2 mm, length $\simeq$ 8.2 mm), silver coated resonant cavity, whose resonance frequency could be mechanically adjusted to a chosen value, $\nu$=48.3 GHz in the present case. The cavity was chosen to work in absorption in the TE$_{011}$ mode. Microwave fields and currents are in the $(a,b)$ planes of the samples, thus avoiding any $c$-axis contribution to the dissipation. The magnetic field was applied aligned with the $c$-axis of the DyBCO monodomain (that is, perpendicular to the sample surface) and perpendicular to the microwave currents (maximum Lorentz force configuration for flux motion). It should be mentioned that, in the present microwave field configuration, the edges and the central part of the samples play a negligible role: the microwave currents vanish at the edges of the cavity walls, and are of maximum intensity on a thin circular band of average diameter 4 mm. The response mainly comes from a circular region of inner diameter 2 mm and outer diameter 6 mm.\\
Measurements of the changes in the quality factor $Q=2\pi\nu W/P$ (where $W$ is the stored energy in the cavity and $P$ the power dissipated on the walls) as a function of the temperature $T$ and the magnetic field $H$ give the changes of the surface resistance $\Delta R_s$ according to the relation:
\begin{equation}
\begin{split}
\Delta R_s (H,T)&=R_{s} (H,T)-R_{s} (0,T)=\\
&=G\left[\frac{1}{Q(H,T)} -  \frac{1}{Q(0,T)}\right]
\label{Reff} 
\end{split}
\end{equation}
\noindent where $Q(H,T)$ is the field-dependent $Q$ factor of the cavity (with the sample mounted), $Q(0,T)$ is the corresponding value measured in zero field, and $G=2\pi\nu W/\frac{1}{2}\int_S|\mathbf{H}_t|^2dS $ is a geometrical factor that can be calculated from the modal expressions of the fields in the cavity. There, $\mathbf{H}_t$ is the tangential magnetic field on the surface $S$. In our case, $G\simeq$~10840 $\Omega$. It should be noted that no calibration of the temperature dependence of the surface resistance of the metal cavity is involved in this procedure: uncertainties in the calculation of geometrical factor only affect the measurements through an overall scale factor \cite{errors} and the $H$ dependence of the change in the surface resistance is not affected by the empty cavity response. Simultaneous measurements of the field-induced shift in the resonant frequency would yield the variations of the surface reactance. However, as detailed in Ref.(\onlinecite{silva}), in the present setup only the changes in $Q$ were measured, being the frequency shift below our sensitivity. We can thus report the field dependence of the surface resistance, but not the surface reactance.
\section{Microwave data and discussion}
\begin{figure}[htb]
\begin{center}
\includegraphics [width=6cm]{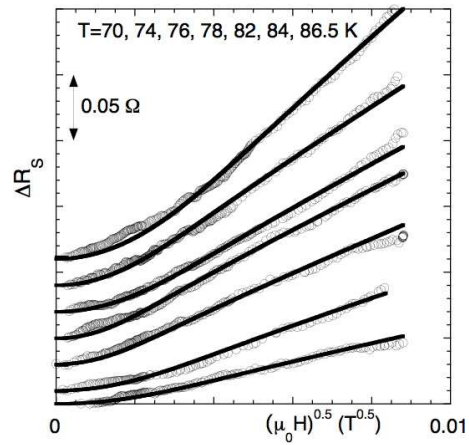}
\caption{Field dependence of the surface resistance change $\Delta R_s$ at $\nu=$ 48.3 GHz as a function of the square root of the applied field, $\sqrt{\mu_0 H}$, at various temperatures. Empty circles: experimental data (50\% of the data is shown to avoid crowding). Continuous lines are two-parameter fits obtained by means of Eq.(\ref{Rs}); the curves are systematically displaced for readability; the upper curve corresponds to the highest temperature. It is seen that the behavior is not a simple square-root.}
\end{center}
\label{rampesqrt}
\end{figure}

\begin{figure}[htbp]
\begin{center}
\includegraphics [width=6cm]{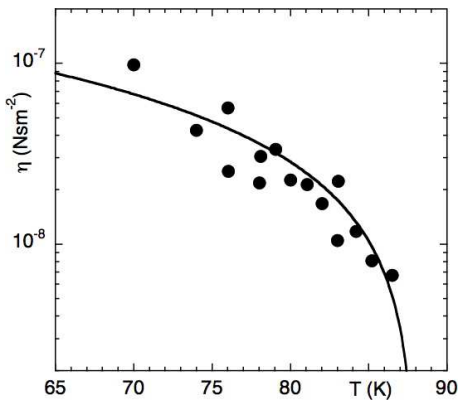}
\caption{Vortex viscosity of the DyBCO monodomain, together with the phenomenological law $\eta(T)=\eta_0(1-t^2)/(1+t^2)$, \cite{golos} with $t=T/T_c$.}
\end{center}
\label{eta}
\end{figure}

\begin{figure}[htbp]
\begin{center}
\includegraphics [width=6cm]{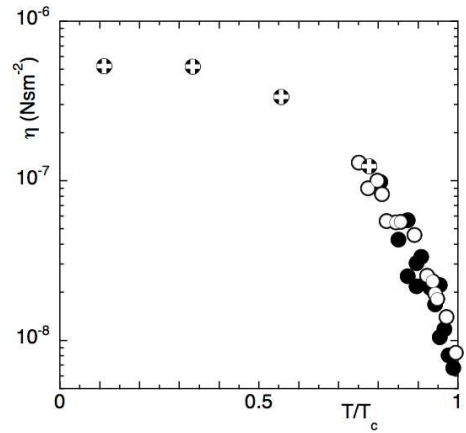}
\caption{Fluxon viscosity dependence versus reduced temperature $T/T_c$, for a set of  123 materials measured at similar frequencies: the DyBCO monodomain of this work, full circles, a SmBCO thin film, \cite{silvaPhC04} open circles, a YBCO single crystal, \cite{tsuchiyaPRB01} crossed circles; a universal behavior  is evident, showing that the electronic state inside the vortex cores is the same in all the materials.}
\end{center}
\label{compare}
\end{figure}

\begin{figure}[htbp]
\begin{center}
\includegraphics [width=6cm]{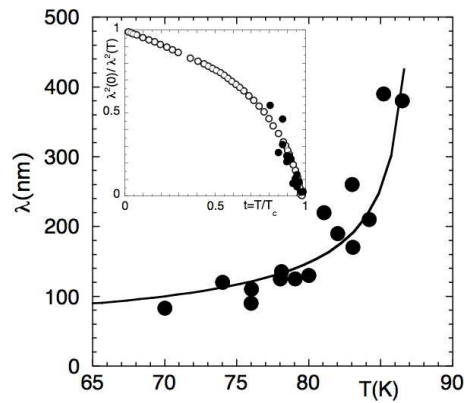}
\caption{Main panel: penetration depth temperature dependence in a DyBCO monodomain (full circles) and two-fluid law given by Eq.\ref{eqlambda} with $n=$2 (d-wave case) and $\lambda_0=$ 65 nm. Inset: comparison of our values of $\lambda$ (full circles, fitting parameters of the magnetic dissipation) and data taken in YBCO single crystals \cite{hosseini} (empty circles), plotted as $\lambda^2(0)/\lambda^2(T)$ vs. $t=T/T_c$. Our data superimpose to the data on single crystals using $\lambda(0)=$ 165 nm.}
\end{center}
\label{lambda}
\end{figure}

We present in the following the main features of the measured microwave response.
A set of illustrative results is reported in Fig.\ref{rampesqrt}, at temperatures ranging from $T/T_c$ = 0.98 till 0.80 and as a function of the square root of the applied field, $\sqrt{\mu_0 H}$. The surface resistance is seen to increase with the applied magnetic field, with downward curvature, and the overall amplitude increases with temperature. Approaching $T_c$ the amplitude decreases, due to the condensate reduction. $\Delta R_s$ is not proportional neither to the applied field, nor to the square-root of the field. As discussed in the following, this is fully consistent with a vortex motion-related dissipation, reflecting the crossover in the screening from superfluid-dominated (at low fields) to flux-flow (effective skin depth) dominated regime.\\
In general, the field induced dissipation at microwave fields can be ascribed to two broad classes of phenomena: fluxon motion and weak-links. When weak links dominate the response, an applied magnetic field of a few tens of mT can easily be strong enough to reduce to zero the phase correlation between different grains. This phenomenon is routinely observed in sintered pellets, \cite{marconPRB89} and can be clearly identified by its very steep field dependence, followed by a saturation when all weak links loose phase correlation (nearly-step-like response to a dc magnetic field). The sample behaves then as several disconnected superconducting islands. Other physical mechanisms for the dephasing, such as heating, \cite{wosik} lead to the same qualitative behavior in $\Delta R_s(H)$: a steep initial rise, followed by a saturation. Thus, it can be safely stated that the presence of weak links can be clearly revealed by the microwave dissipation in a magnetic field depending on the qualitative shape of the response.\\
The ubiquitous process is the motion of flux lines: when a magnetic field $H$ is applied below $T_{c}$, and a transport current is injected in the sample, flux lines move and dissipation is observed in many properties, in particular in electrical and also thermal transport processes. \cite{MA} In the present case the transport current is the microwave current. The problem of the coupled response of vortices, superfluid and quasiparticles to a microwave field has been theoretically solved by Coffey and Clem (CC), \cite{cc} in the approximation of almost uniform vortex density, well above $H_{c1}$. The resulting expression for the surface resistance of a type-II superconductor is quite complex, since it contains all the details related to flux-creep and to the interplay between the normal-fluid penetration depth, the London penetration depth, and the vortex motion complex penetration depth. However, the theoretical expression can be cast in a very simple form if: \\
(i) the London penetration depth is much shorter than the normal fluid penetration depth, which is always true apart very close to $T_{c}$.\\
(ii) the effect of flux-creep can be neglected (that is the "creep factor" is taken to be 0).\\
In our case, each vortex oscillates over time scales much shorter than the time needed for the escape from a potential well and the relaxation time for elastic restoring force by the pinning potential, so that the effects of creep and pinning become irrelevant (even if they can have strong effects on the dc properties). We will see in the following that this assumption finds a strong consistency check in the quantitative analysis of our data.\\
Under constraints {\it (i)} and {\it (ii)}, one obtains \cite{marconPRB91, silvaGd} an expression for the surface resistance analogous to the the well-known model for the vortex motion resistivity found by Gittleman and Rosenblum; \cite{gittle} the change in surface resistance reads:
\begin{equation}
\Delta R_s (B,T)= \frac{\omega\mu_0\lambda}{\sqrt{2}}
\sqrt{-1+
\sqrt{
1+\left(\frac{1}{\omega\mu_0\lambda^2}\frac{\Phi_{0} B}{\eta}\right)^2
}
}
\label{Rs} 
\end{equation}
where $B\simeq \mu_{0} H$ (London limit), and $\eta$ is the so-called vortex viscosity. Here we have neglected the effect of pinning due to our very high measuring frequency.
Notice that this expression, Eq.(\ref{Rs}), like the entire CC theory, is obtained in the mean-field validity range, so that it does not include fluctuations. We believe that this is a more stringent constraint than the point \textit{(i)} above in limiting the applicability of the theory, in particular very close to $T_{c}$.\\
Finally we observe that the theoretical expression Eq.(\ref{Rs}) predicts a crossover from the linear behavior at low fields, 
\begin{equation}
\Delta R_s =\frac{1}{\lambda}\frac{\Phi_{0} B}{\eta},
\label{Rslow} 
\end{equation}
when the material has a (real) flux flow resistivity $\rho_{ff}=\frac{\Phi_{0} B}{\eta}$ and screening is dictated by superfluid over the London penetration depth $\lambda$, to the square-root dependence at higher fields,
\begin{equation}
 \Delta R_s=\sqrt{\frac{\mu_0 \omega}{2}\frac{\Phi_{0} B}{\eta}},
 \label{Rshigh} 
\end{equation}
typical of the skin-depth regime, where screening is given by  
\begin{equation}
\delta_{ff}=\sqrt{\frac{2\rho_{ff}}{\mu_0\omega}}. 
 \end{equation}
Fitting of the data in the crossover region, as in our case (see Figure \ref{rampesqrt}), allows for a true two-parameter (i.e. $\lambda$, $\eta$) fit. The fits are reported as continuous lines in Figure \ref{rampesqrt}. As can be seen, Eq.(\ref{Rs}) represents an excellent description of our data. Correspondingly, we obtain values of the fluxon viscosity $\eta$ (Figs. \ref{eta} and \ref{compare}) and the penetration depth $\lambda$ (Fig. \ref{lambda}) as a function of the temperature. \\
It should be noted that the evaluation of the fluxon viscosity brings considerable information about the nature of the flux lines involved in the motion, and on the nature of the electronic properties in the vortex cores. In fact, in well-connected, crystalline-quality materials flux lines nucleate as Abrikosov vortices (AV), with a well-defined core where quasiparticle states are excited, and the vortex viscosity can be phenomenologically expressed using the Bardeen-Stephen model for flux flow, yielding \cite{golos} $\eta_0(1-t^2)/(1+t^2)$, with $t=T/T_c$. In presence of small cracks, reasonably extended point defects and large angle grain boundaries, also Josephson vortices (JV) and Abrikosov-Josephson vortices \cite{gurevichPRB92,SSMA} (AJV) can nucleate. \\
The corresponding vortex viscosity is however dramatically different, \cite{halbritterJS95} for the JV have no core in the common sense, and nucleate in a region where the density of states is vanishingly small. AJV have a somewhat intermediate behavior, depending on the grain boundary angle. As a consequence, besides comparing the data to some theoretical equation, it makes much sense to compare the absolute values of the vortex viscosity as measured in our single-domains with corresponding data measured in single crystals, since in the latter only AV should be present. Due to the lack of measurements of the vortex viscosity in DyBCO single crystals, in the following we compare our results to data obtained in YBCO. \cite{tsuchiyaPRB01} In addition, we include data taken in a SmBCO thin film \cite{silvaPhC04} to give an overview of the response of 123 materials.\\
Figure \ref{eta} reports the viscosity data, obtained from the fits in Fig. \ref{rampesqrt} for our DyBCO monodomain. The phenomelogical expression is in good agreement with the data, with $\eta_0=$3 10$^{-7}$Nsm$^{-2}$.\\
Comparison with two other materials of the 123 family is accomplished in Fig.\ref{compare}. We first note that the magnitude of the viscosity in DyBCO compares excellently (on an absolute scale) to the data taken in YBCO. This is a significant point, since it involves that the monodomain here under study behaves basically as a single crystal. Secondly, and most important, the plot of $\eta$ vs. $t$ reveals a universal behavior, strengthening the indication for identical electronic states and dissipation mechanisms.\\
The fact that the electronic state in DyBCO is the same as in YBCO and related compounds is fortified by the analysis of the penetration depth. Figure \ref{lambda} reports the subsequent data obtained from the fit for the temperature dependent real part of the penetration depth in the DyBCO monodomain, which shows a considerable increase and divergence near $T_c$. The present data compares excellently to the theory based on the two fluid model, \cite{ma63} i.e.
\begin{equation}
\lambda(T) = \lambda_0 / \sqrt{1-(T/T_c)^n}
\label{eqlambda}
\end{equation}
where $n$ = 4 or 2 depending whether the gap is $s-$ or $d-$ wave  type symmetry, and $\lambda_0$ represents the zero-temperature expectation for the penetration depth within the simple two-fluid fit. In the relatively small temperature range here explored the data cannot discriminate between the two fits, only  the value of $\lambda_0$ being affected. We find  $\lambda_0$= 60 nm and 80 nm in the $d-$ and $s-$ wave case, respectively. It should be mentioned that $\lambda_0$ is not a correct evaluation of the true zero-temperature London penetration depth, due to the fact that it extrapolates to zero temperature from data taken in the high-$T$ region. In fact, it is known \cite{waldramPRB97} that the temperature dependence of $\lambda(T)$ in $d-$wave cuprates can hardly be described in the full temperature range by a single expression of the elementary form as Eq.(\ref{eqlambda}). An instructive insight can be gained by direct comparison with data taken in single crystals, as reported in the inset of Fig.\ref{lambda}. There, $\lambda$ from our fits superimpose to the data taken on YBCO single crystals, represented in Ref.(\onlinecite{hosseini}) as $\left[\lambda(0)/\lambda\right]^2$, using in this case $\lambda(0)=$ 165 nm as a scale factor, in close agreement with expected values for 123 thin films.\\
Finally, we comment on the robustness of the assumption of the free flux flow limit of the vortex complex response, that is neglecting possible effects of pinning. When dealing with microwave response, one needs to recall that only extremely small vortex oscillations [of order of 0.1 nm, see Ref.(\onlinecite{TomaschPRB88})] are imposed by a microwave field. The higher the working frequency, the smaller the oscillation amplitude, and the dynamics can be considered as that of independent motion of single vortices (no collective effects). This picture eventually breaks down when vortices become very closely packed, and even such small oscillations become a sizeable perturbation of the static configuration. We argue that this regime is confined to high fields and/or to temperatures very close to $T_c$, and is not reached in our field and temperature range. Indeed, low frequency (9.5 GHz) microwave measurements in DyBCO films gave a ``peak effect" \cite{bhangaleJAP01} reminiscent of the ordered-disordered vortex lattice crossover. However, it has been shown \cite{bhangaleJAP01} that decreasing the frequency, as to 4.75 GHz, would lead to a substantial reduction of this effect. This behavior, apart from being sample-dependent, can be taken into account by an operating frequency close to the depinning frequency. In fact, a depinning frequency $\nu_p$ of the order of 10 GHz has been estimated, \cite{bhangaleJAP01} consistently with observations in YBCO.\cite{golos} Our experimental results are consistent with this estimate, since we do not observe any trace of the peak effect. In that case, taking $\nu_p\approx$10 GHz, since the correction factor to the surface resistance is $\sim 1/\left[1+(\nu_p/\nu)^2\right]$, \cite{golos,cc} we would need to correct our data for the viscosity at most by $\sim$4\%, which leaves all of the results of this paper rather unaffected.

\section{Conclusion}

Before concluding let us observe that the behavior of  the viscosity and penetration depth allows us to point out the relevance of comparing quantities in reduced units: indeed in Fig.(\ref{lambda}), inset, and in Fig.(\ref{compare}) we show in reduced temperature units, i.e. as a function of  $T/T_c$, the behavior of the fluxon viscosity and the penetration depth for a variety of materials of the 123-HTcS family, some measured in Ref.(\onlinecite{silvaPhC04}). In so doing a $universal$ law is  easily observed.
Whence, we can obtain definite conclusions on the field and temperature dependence of the surface resistance of this DyBCO monodomain. We have obtained the fluxon viscosity and penetration depth as a function of temperature from the surface resistance at 48.3 GHz. We have found that (1) the dissipation originates most probably from Abrikosov vortices, whose cores have an electronic state identical to the one inferred from measurements in YBCO; (2) the viscosity obeys the well-known phenomenonological law $\eta\propto(1-t^2)/(1+t^2)$, where $t=T/T_c$; (3) the penetration depth obeys a two-fluid law in the vicinity of $T_c$, as is the case for other 123-crystals.
Moreover it seems that we have grown large DyBCO samples that exhibit single-crystal-like behavior with respect to the application of a magnetic field, indicating an interesting step toward potential applications. The similarity of the vortex parameters of DyBCO monodomains (measured here) to those known on other materials allows the design of applications by inferring the properties of DyBCO from the vast body of measurements in e.g. YBCO.

\section*{\textsc{Acknowledgments}}
Useful help by Dr. Denis in the preliminary stage of this work is kindly acknowledged. This work has been partially supported by an exchange program under the Scientific Cooperation Program between Italy and the Communaut\' e  Fran\c{c}aise de Belgique (CGRI-FNRS) (2007-2008).


\end{document}